\documentclass[fleqn,usenatbib]{mnras}

\usepackage{newtxtext,newtxmath}

\usepackage[T1]{fontenc}
\usepackage{ae,aecompl}

\usepackage{graphicx}	
\usepackage{amsmath}	
\usepackage{amssymb}	

\title[The time delay of JVAS~B1030+074]{The time delay of JVAS~B1030+074 from VLA polarization monitoring}

\author[A. D. Biggs et al.]{
  A.~D.~Biggs$^{1}$\thanks{E--mail: abiggs@eso.org}
  \\
  $^{1}$European Southern Observatory, Karl Schwarzschild Stra{\ss}e 2, D-85748 Garching, Germany
}

\date{Accepted XXX. Received YYY; in original form ZZZ}

\pubyear{2018}

\begin{document}
\label{firstpage}
\pagerange{\pageref{firstpage}--\pageref{lastpage}}
\maketitle

\begin{abstract}
  We have analysed archival VLA 8.4-GHz monitoring data of the gravitational lens system JVAS~B1030+074 with the goal of determining the time delay between the two lensed images via the polarization variability. In contrast to the previously published total intensity variations, we detect correlated variability in polarized flux density, percentage polarization and polarization position angle. The latter includes a fast ($<$5~d) 90-degree rotation event. Our best estimate of the time delay is $146\pm6$~d (1~$\sigma$), considerably longer than that predicted by the lens model presented in the discovery paper. Additional model constraints will be needed before this system can be used to measure $H_0$, for example through a detection of the lensed source's VLBI jet in image B. No time delay is visible in total flux density and this is partially due to much greater scatter in the image B measurements. This must be due to a propagation effect as the radio waves pass through the ISM of the lensing galaxies or the Galaxy.
\end{abstract}

\begin{keywords}
  quasars: individual: JVAS~B1030+074 -- gravitational lensing: strong -- cosmology: observations -- galaxies: ISM
\end{keywords}



\section{Introduction}
\label{sec:intro}

The measurement of gravitational lens time delays offers a single-step determination of the expansion rate of the Universe, $H_0$, at cosmological distances, independent of any intermediate ``distance-ladder'' calibrations \citep{refsdal64}. Given that current determinations of $H_0$ based on observations of relatively local Cepheid-calibrated supernovae \citep{riess16} and the Cosmic Microwave Background \citep{planck16} show signs of inconsistency at the 3-$\sigma$ level, it is perhaps more important than ever that additional methods, such as that offered by lens time delays, be pursued in an effort to try and resolve the potential discrepancy.

The lens system B1030+074 \citep[][hereafter X98]{x98} is one of six gravitational lens systems discovered as part of the Jodrell Bank/VLA Astrometric Survey \citep[JVAS --][]{king99}. It consists of two images (A and B) of a $z=1.535$ radio-loud quasar lensed by a spiral galaxy at $z=0.599$ \citep{fassnacht98}. The lensing galaxy is extended westwards of its nucleus and it is unclear if this is substructure in the galaxy or a separate galaxy \citep*{jackson00,lehar00}. The image separation is 1.6~arcsec and lens modelling performed by X98 predicts a time delay of 112~d for $H_0 = 70$~km\,s$^{-1}$\,Mpc$^{-1}$.

A number of attempts have been made to determine the time delay, the most useful of which utilised the Very Large Array (VLA) observing at a frequency of 8.4~GHz. A first season of monitoring in 1998 detected moderate changes in the total flux density of image A, but no corresponding features were visible in B \citep{x05}. Further monitoring was carried out in 2000/2001 \citep{rumbaugh15} but the total-flux-density variability again proved insufficient to measure a delay.

Ignored by practically all radio campaigns to date has been the potential of monitoring variations in the polarization properties of the source i.e.\ polarized flux density, percentage polarization and polarization position angle (PPA). A notable exception is JVAS~B0218+357 for which polarization monitoring has been an essential part of determining the time delay \citep{corbett96,biggs99,biggs18}. However, B0218+357 is particularly highly polarized (approaching 10~per~cent at high frequencies) and also bright, the total flux density of each image exceeding 100~mJy.

In general, radio cores are less polarized than extended emission and few radio lenses are as bright as B0218+357. However, polarization monitoring should be of great interest to time-delay studies as the magnitude of variability tends to be greater than that seen in total intensity and the timescale of variability shorter \citep[e.g.][]{saikia88}. In addition, the PPA is occasionally observed to rapidly rotate \citep{lyutikov17}, an event which should be observable for even a weakly polarized source.

We can find no mention in the literature of whether the lensed images of B1030+074 are polarized or not, but our own analysis of archival VLA data revealed polarization at the 1--2~per~cent level. We have therefore reanalysed the VLA 8.4-GHz monitoring data from 1998 and present here the resultant total-intensity and polarization variability curves. We do not include the \citet{rumbaugh15} data as much less time was spent observing the lens during this campaign and the signal-to-noise ratio (SNR) of the polarization measurements is too low to reliably detect image B.

\section{Observations and data reduction}
\label{sec:obs}

The data presented in this paper correspond to VLA project AX004 and consist of 46 observations of B1030+074 between 1998 February 19 and 1998 October 11, an average sampling rate of one epoch every 5.1~d. The campaign began during the move to the VLA's A configuration and ended with the antennas in their B configuration. Data were taken at 8.4~GHz using the standard continuum mode of two 50-MHz-wide subbands with the correlator producing all four combinations of the right and left circularly polarized signals (RR, LL, RL and LR) necessary for polarization measurements. At this frequency, the angular resolution is approximately 0.2 and 0.6~arcsec in the A and B configurations, significantly less than the separation between the two lensed images of 1.6~arcsec.

The observing sequence during each epoch was the same. As well as the lens, three calibrators were included: 1038+064 (a core-dominated source with a faint jet), 1005+07 (3C~237) and 1117+146 \citep*{bondi98}. The last two are both Compact Symmetric Objects (CSOs), a class of source which is known to vary very little due to being dominated by emission from extended radio lobes \citep{fassnacht01}. B1030+074 was observed for 26~min at all epochs, except for one which was terminated early due to a lightning strike.

The data were calibrated using NRAO's Astronomical Image Processing System (\textsc{aips}). After calibration of the gain-elevation response of each antenna using \textit{a priori} calibration curves, models for all three calibrators were created using mapping and self-calibration. Our chosen flux density reference was 1117+146 and we estimated its flux to be 615~mJy bootstrapped from a value of 1.03~Jy for 3C~237 \citep{mantovani09}. Each epoch was then calibrated using standard procedures, the gain solutions found for 1038+064 being interpolated onto B1030+074 and the phases of the lens calibrated with a model of the source.

Polarization calibration (leakage and PPA) proved more difficult as no standard polarization calibrators were included in the observing schedule. The polarization leakage (or D-terms) can either be calibrated using observations of sources of unknown polarization over a wide range of parallactic angle or by using a single scan of an unpolarized source. The first was not possible given the relatively short observation time of each epoch ($\sim$1~h) and the proximity of all calibrators to the lens, but the second approach seemed worthy of investigation given that CSOs tend to be unpolarized \citep{peck00}. We confirmed that this is the case for 1117+146 by calibrating the polarization leakage using a number of epochs for which additional calibrators associated with monitoring of other lens systems were available and combining these into a single dataset. We estimate a polarized flux density of 0.3~mJy\,beam$^{-1}$ or $<$0.05~per~cent.

Another complication is caused by the fact that 1117+146 is significantly resolved in A configuration. As such, the standard mode (`\textsc{appr}') used with the \textsc{aips} leakage-calibration task \textsc{pcal} gives poor results i.e.\ the solutions differed significantly from those found using the calibrators associated with the other lens systems monitored as part of AX004 and polarization maps contained ugly residuals. Fortunately, \textsc{pcal} provides an additional mode (`\textsc{rapr}') for use with resolved calibrators and this works well -- consistent leakage terms are found for all antennas over the full duration of the monitoring and no residuals are present in polarization maps of B1030+074. We note that for this mode to work, it is necessary to first remove the parallactic angle from the phases of each antenna using \textsc{clcor}. Failure to include this step produces inferior results.

Calibrating the PPA is also problematic. Although 1038+064 is significantly polarized, it is a point source and thus probably variable. The polarization of 3C~237, on the other hand, is dominated by the western lobe and therefore unlikely to vary, but its polarized flux is weaker, having a comparable polarized flux density to that of image A of B1030+074. We have therefore followed a two-stage process. We firstly aligned the two IFs at each epoch by measuring the PPA of 1038+064 in each IF separately -- this correction was applied to all sources in \textsc{aips}. The PPA of 3C~237 was then measured from combined-IF maps and used to correct that of each image of B1030+074.

The data for the variability curves were derived by model fitting to the $u,v$ data. This was performed using \textsc{difmap} \citep{shepherd97} using two delta components at the known positions of images A and B. We first fit to the Stokes \textit{I} visibility data in conjunction with phase-only self-calibration. After the residual phases had been corrected, we also modelfit to the Stokes \textit{Q}, \textit{U} and \textit{V} data and form the polarized flux density, percentage polarization and PPA measurements from the usual combination of \textit{Q}, \textit{U} and \textit{I}. The chi-squareds of the fits are close to unity, except for a small number of epochs which are affected by bad weather. If an amplitude-calibration step is incorporated into the modelfitting we find $\chi^2 \le 1.21$ for all epochs and Stokes parameters, with only very minor differences in the fitted flux densities. Therefore, we are satisfied that there are no major problems with the data and exclude none of the 46 epochs from the time-delay analysis.

The uncertainties on the flux density measurements are calculated through the quadrature sum of a thermal noise term (measured from the residual maps) and a 0.5~per~cent flux-scale error. In addition, the polarized flux density error includes a term to account for residual errors in the polarization leakage calibration and we have assumed that these are proportional to the Stokes $I$ flux density ($d\,I$) with $d = 0.1$~per~cent (R.~Perley, private communication). The signal-to-noise ratio of the polarized flux densities is often very low (SNR$<$3) and the measured values are therefore subject to positive Ricean bias. We correct for this using the approach of \citet{wardle74} although the corrections are very small and have no effect on the derived time delay. The PPA measurements are not subject to bias, but their probability distribution becomes increasingly non-Gaussian at low SNR with significant wings. \citeauthor{wardle74} introduce two approximations to a one-sigma error and we have conservatively used that which produces the largest uncertainties for our data (their equation A6). No flux-scale error is included in the percentage polarization uncertainties as this cancels out in the ratio of polarized to total flux density.

\section{Variability curves}
\label{sec:results}

\begin{figure*}
\begin{center}
\includegraphics[scale=0.27]{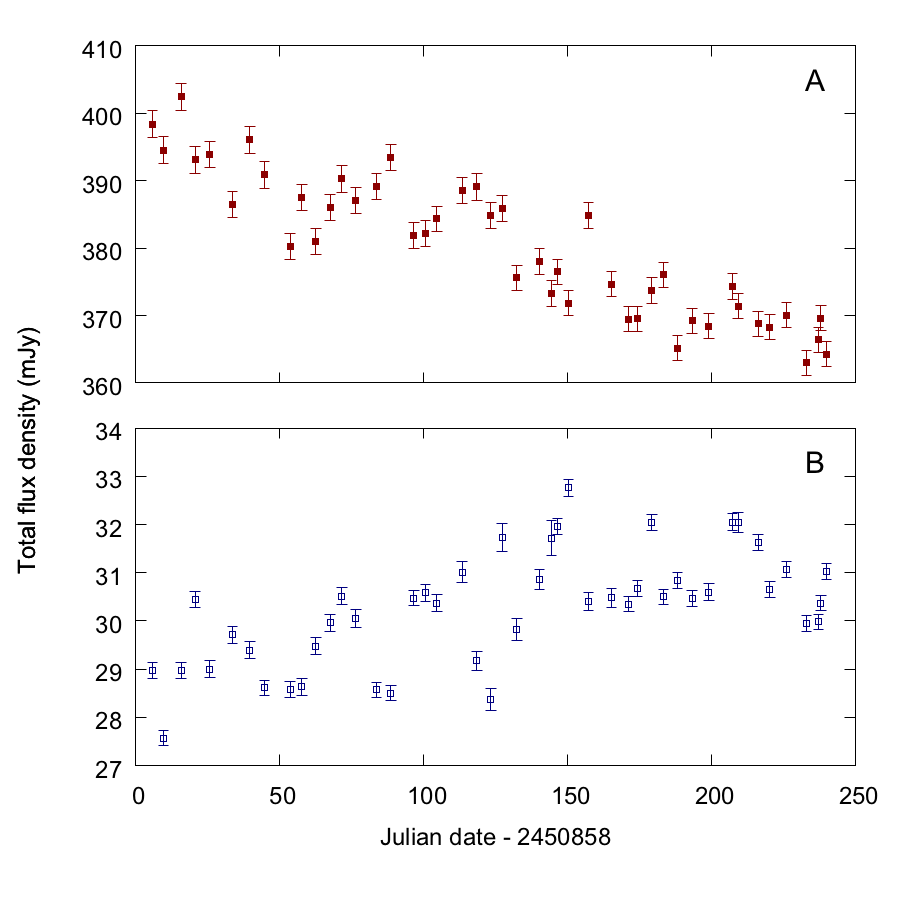}
\includegraphics[scale=0.27]{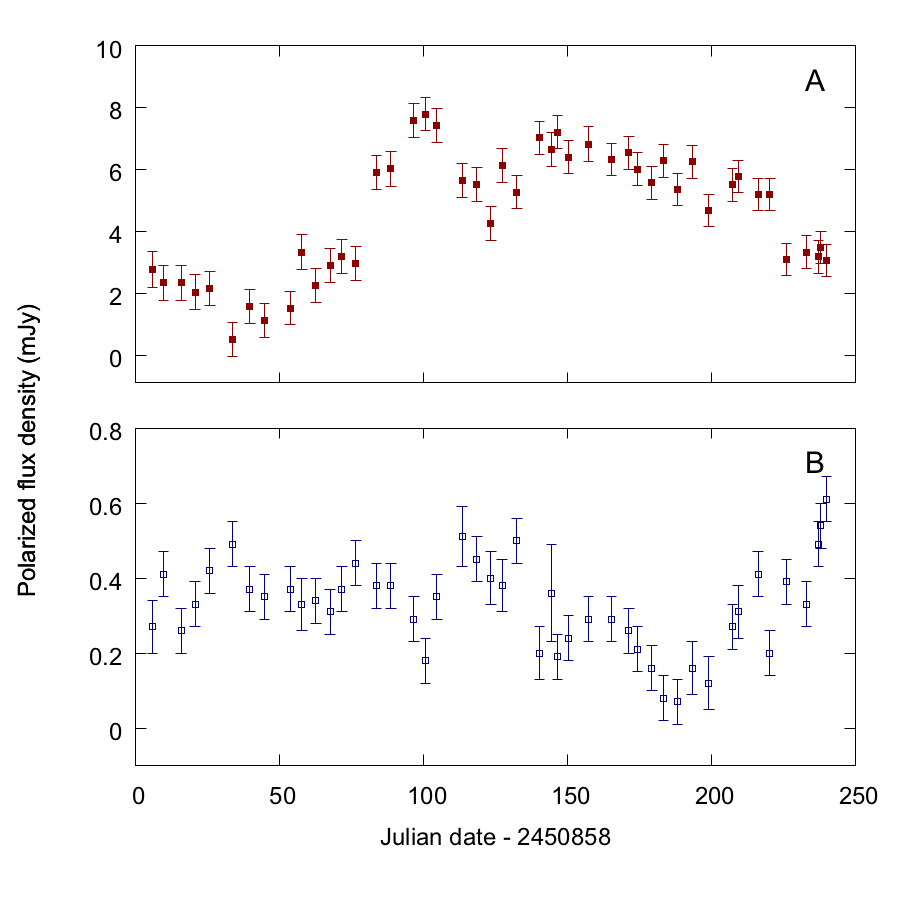}
\includegraphics[scale=0.27]{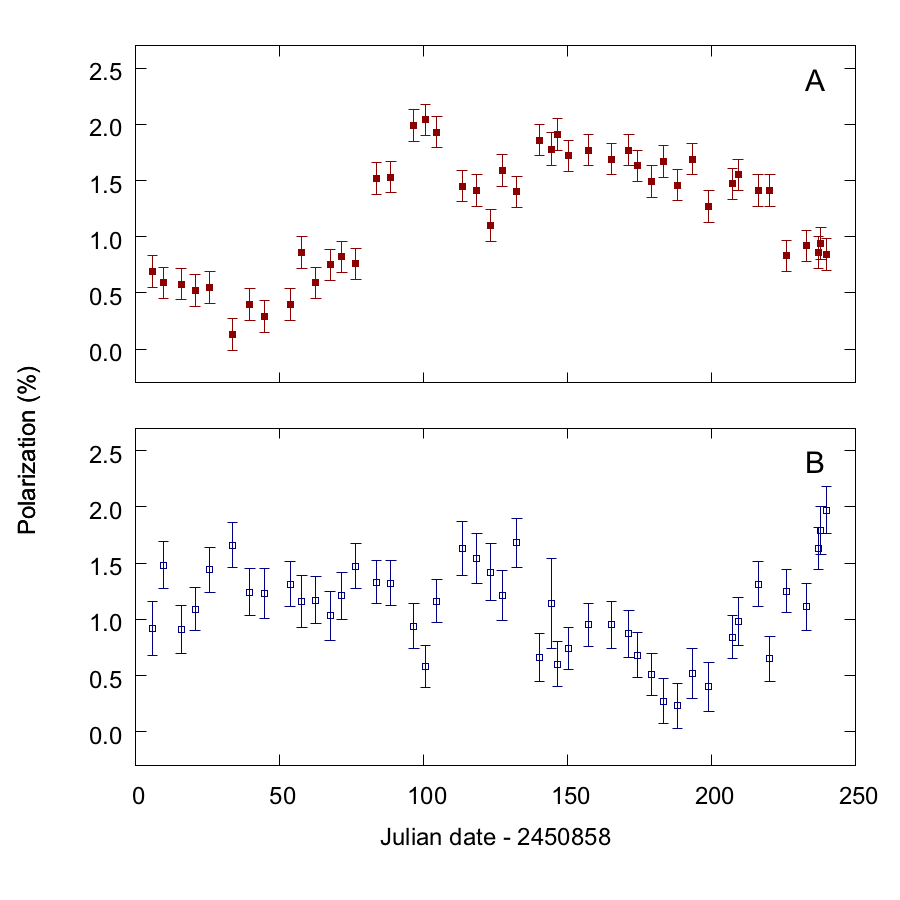}
\includegraphics[scale=0.27]{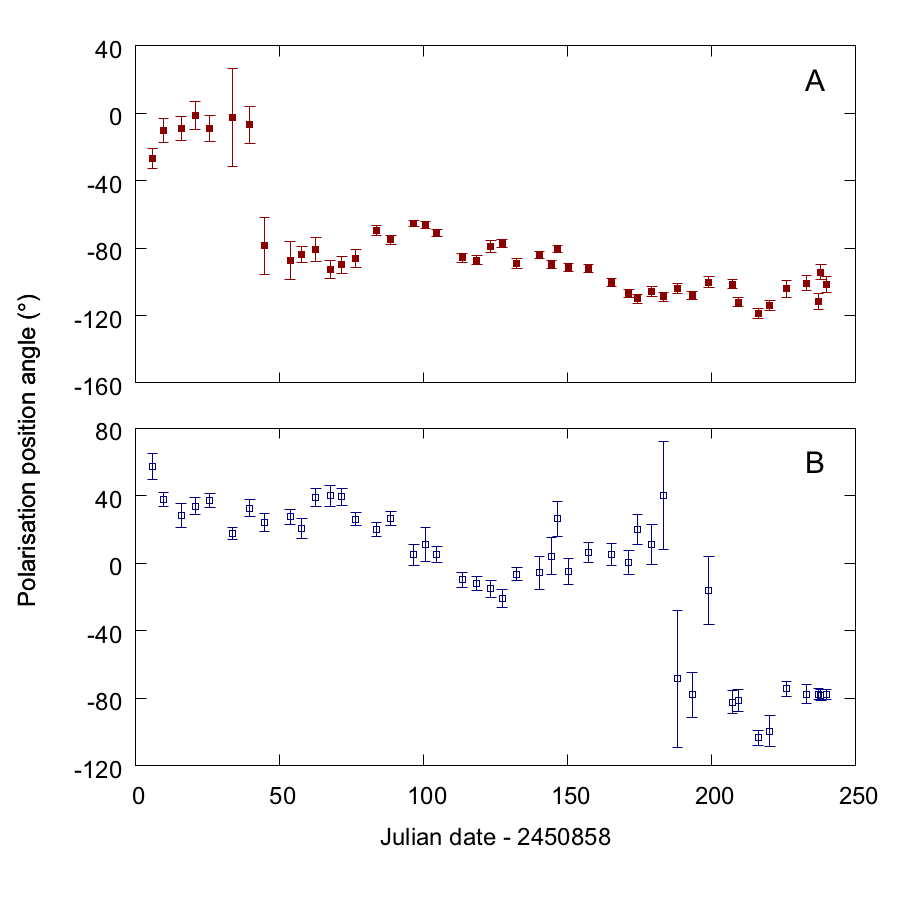}
\caption{VLA 8.4-GHz variability curves of B1030+074. Top-left: total flux density, top-right: polarized flux density, bottom-left: percentage polarization and bottom-right: polarization position angle (PPA). Image A is the top panel in each case.}
\label{fig:axlc}
\end{center}
\end{figure*}

The 8.4-GHz variability curves are shown in Fig.~\ref{fig:axlc}. The total flux density variations are very similar to those shown by X98, with no obvious features common to both images that might indicate the time delay. Image A declines linearly by about 10~per~cent, whilst image B increases by a similar amount. There is, however, a hint that image B is beginning to decline towards the end of the monitoring, as it must if the time delay is not longer than the predicted value (112~d) by more than a factor of two. Identifying any turnover in image B is made more difficult due to the increased scatter compared to A -- the residuals around a straight-line fit are twice as large as for image A, despite image B being fainter by more than a factor of ten.

Greater variability is present in the polarization data. Towards the beginning of the monitoring there is a drop in image A's polarized flux density to nearly zero, followed by an approximately linear increase to $\sim$8~mJy. Such inflection points are crucial for measuring a time delay and the same feature appears to be present in image B around 150~d later. The percentage-polarization variations are almost identical to those seen in polarized flux density and the maximum detected is $\sim$2~per~cent, typical for a flat-spectrum quasar.

The polarization minimum is associated with a sudden, large rotation in the PPA. This is seen quite clearly in both images and the average of the rotations seen in each image is equal to $-$90.1\degr. Rotation events with a magnitude of 90\degr\ are commonly associated with polarization minima and are thought to correspond to changes in the opacity of the jet \citep{wardle18}. Regardless of the cause of the rotation, it is so abrupt (the swing itself does not seem to be fully sampled despite an average separation of only 5~d) that it should provide a good constraint on the time delay.

On the basis of a by-eye inspection alone, it is therefore clear that the time delay is present in all polarization variability curves and we henceforth exclude the total flux density data from the time-delay analysis. We also do not consider the polarized flux density any further as this is practically identical to the percentage polarization data and the latter should in principle be more robust as they are not affected by errors in the flux scale.

A sample of the data format is shown in Table~\ref{tab:data} for percentage polarization.

\begin{table*}
  \centering
  \caption{The first five epochs of the percentage polarization ($p$) variability curves. All four datasets (total flux density, polarized flux density, percentage polarization and PPA) are available in full online. The last four columns give the $\chi^2$ of the modelfit to the visibility data and the noise ($\sigma$) in a naturally-weighted residual map for both the Stokes \textit{Q} and \textit{U} data. Only one $\chi^2$ and rms is reported for Stokes $I$.}
  \label{tab:data}
  \begin{tabular}{cccccccccc} \\ \hline
Julian date & Hour angle (h) & $p_{\mathrm{A}}$ (\%) & $\Delta p_{\mathrm{A}}$ (\%) & $p_{\mathrm{B}}$ (\%) & $\Delta p_{\mathrm{B}}$ (\%) & $\chi^2_Q$ & $\chi^2_U$ & $\sigma_Q$ (mJy\,beam$^{-1}$) & $\sigma_U$ (mJy\,beam$^{-1}$) \\ \hline
2450863.929 &  2.508 & 0.69 & 0.14 & 0.92 & 0.24 & 1.06 & 1.07 & 0.058 & 0.045 \\
2450867.772 & -1.003 & 0.59 & 0.14 & 1.48 & 0.21 & 1.06 & 1.07 & 0.045 & 0.041 \\
2450873.818 &  0.503 & 0.58 & 0.14 & 0.91 & 0.21 & 1.07 & 1.07 & 0.041 & 0.047 \\
2450878.763 & -0.497 & 0.52 & 0.14 & 1.09 & 0.19 & 1.06 & 1.07 & 0.038 & 0.041 \\
2450883.770 &  0.002 & 0.55 & 0.14 & 1.44 & 0.20 & 1.07 & 1.07 & 0.035 & 0.046 \\ \hline
  \end{tabular}
\end{table*}

\section{Time-delay analysis}
\label{sec:delay}

To determine the time delay, we have performed a standard analysis using a chi-squared minimum \citep[CSM -- e.g.][]{kundic97} and a cross-correlation technique \citep[CCF -- e.g.][]{white94}. In both cases, one image is shifted by a trial time delay (in steps of 1~d) and each point compared with its equivalent in the other image using interpolation. One notable difference is that the CSM also removes the $y$ offset (depolarization or Faraday rotation) when performing the variability-curve comparison.

\begin{figure}
\begin{center}
\includegraphics[scale=0.27]{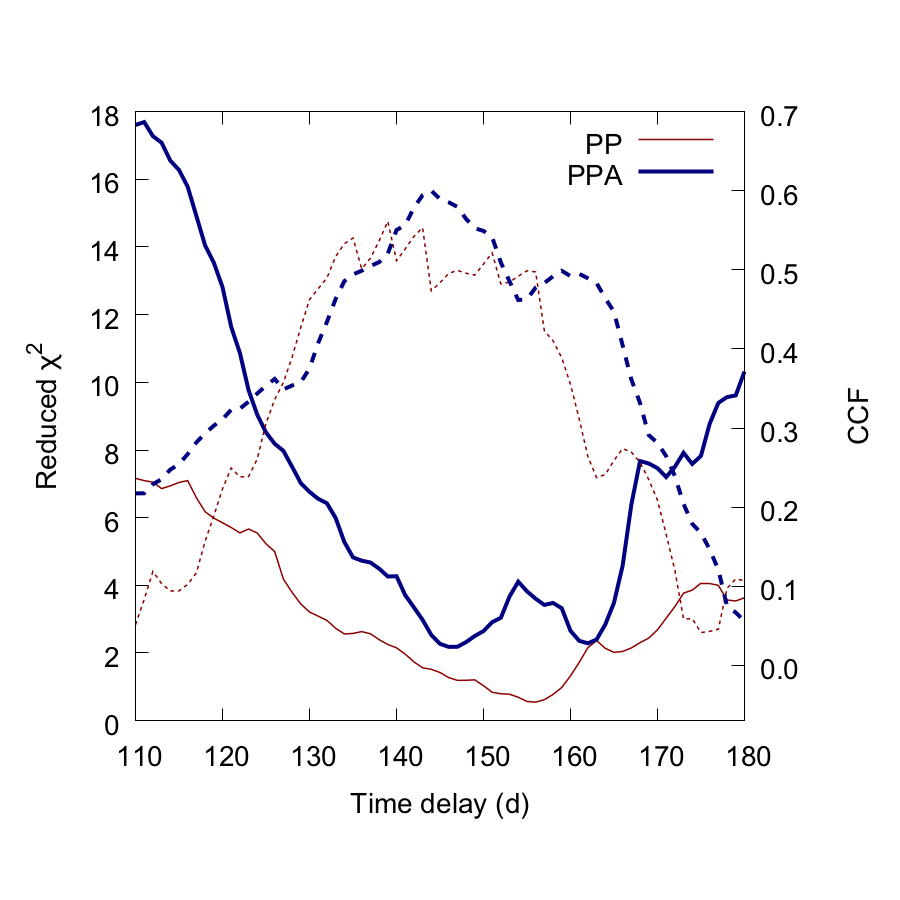}
\caption{Cross-correlation (dashed line) and reduced chi-squared (solid line) as a function of trial time delay for the percentage polarization and PPA variability curves.} 
\label{fig:chiccf}
\end{center}
\end{figure}

The results are shown in Fig.~\ref{fig:chiccf} where it can be seen that the PPA data, as expected, give a more pronounced peak/minimum and very similar results for the CCF and CSM techniques of 144 and 146~d, respectively (Table~\ref{tab:delay}). The percentage polarization data constrain the delay less well and the CCF, for example, has a relatively broad and jagged plateau extending between 134 and 156~d. Table~\ref{tab:delay} also shows the results of applying the dispersion-minimization statistic $D^2_2$ of \citet{pelt94,pelt96} which does not utilize interpolation. As with the CSM technique, this also removes any $y$ offset and both give very similar results.

\begin{table}
  \centering
  \caption{Best-fit time delays for the cross-correlation (CCF), chi-squared minimisation (CSM) and Pelt dispersion techniques. The final column gives the $y$ offset between the two variability curves for the CSM technique: $^{\dagger}$ polarization ratio (A/B), $^{\ddagger}$ Faraday rotation (A$-$B) in degrees.}
  \label{tab:delay}
  \begin{tabular}{ccccc} \\ \hline
    Dataset & \multicolumn{3}{c}{Time delay (d)} & $y$ offset \\
     & CCF & CSM & $D^2_2$ & CSM \\ \hline
    Percentage polarization     & 139 & 156 & 154 & 0.72$^{\dagger}$ \\
    Polarization position angle & 144 & 146 & 146 & 1$^{\ddagger}$ \\ \hline
  \end{tabular}
\end{table}

The minimum chi-squared value of the percentage polarization data is 0.55, indicating a good match between the variations in each image, but also that the errors may have been overestimated. Conversely, the equivalent value for the PPA data is greater than 1 ($\chi^2 = 2.2$). This is at least partly due to Epoch~37 (31 August) for which the PPA of image B seems to be discrepant. This is clearly visible in Fig.~\ref{fig:axlc} as the third epoch after the large PPA rotation and seems to show that the PPA briefly returns to the pre-rotation state. If this is indeed the case, no obvious counterpart to this event is visible in image A, thus leading to an artificially high chi-squared value.

\begin{figure*}
  \begin{center}
    \includegraphics[scale=0.27]{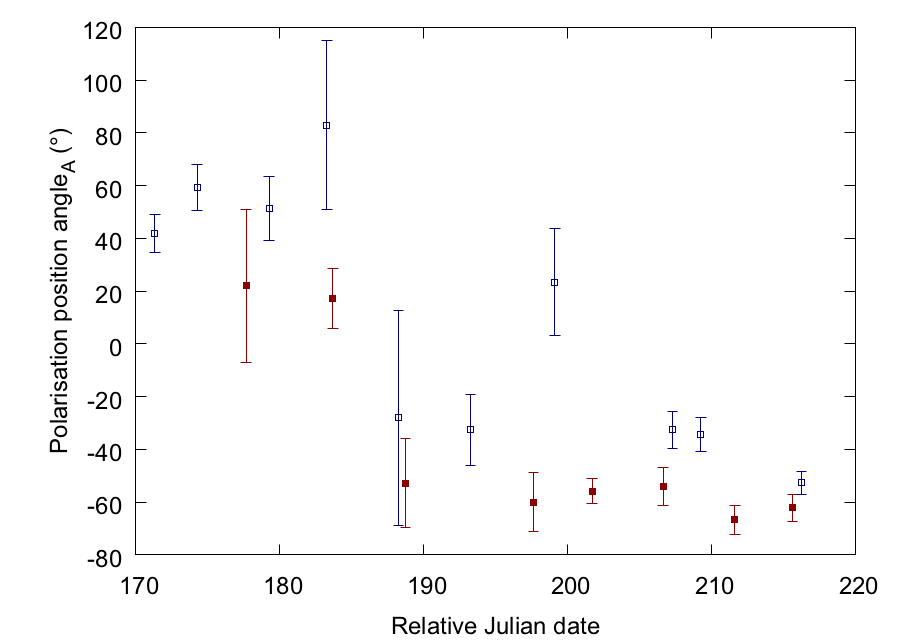}
    \includegraphics[scale=0.27]{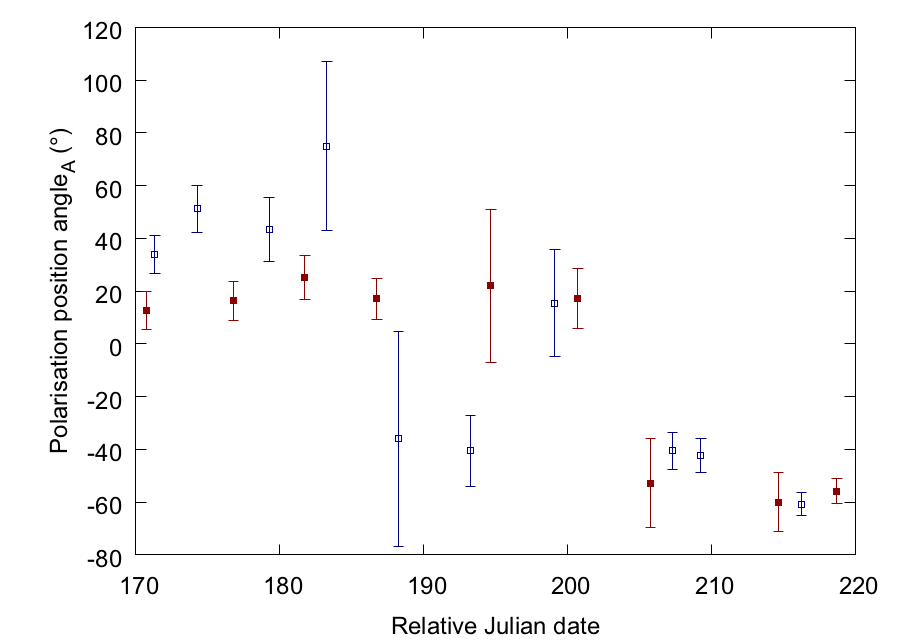}
    \caption{Zooms of the PPA rotation event after shifting A forward by the CCF best-fit time delays of 144~d (left) and 161~d (right). Image A: filled symbols, image B: open symbols. It can be seen that both delays are plausible providing that there are three rapid ($\la$2~d) 90-degree swings.}
    \label{fig:rotevent}
  \end{center}
\end{figure*}

The effect of Epoch~37 is to produce an additional feature in both the CSM and CCF distributions at a higher delay of $\sim$161~d. Although we think that the lower delay ($\sim$145~d) is more likely, we should explore the possibility that the feature corresponding to the higher delay is the correct one. In this scenario, the counterpart to the second rotation event \textit{is} seen and corresponds to the last two epochs in image~A before the sudden PPA swing. An unsampled rotation event still exists and it is now the two earlier epochs (35 and 36) which have no counterpart in image~A. Both scenarios are illustrated in Fig.~\ref{fig:rotevent} where we show times around the rotation event(s) for the most likely CCF delays of 144 and 161~d. In each case it can be seen that the unsampled features can be arranged such they lie inside gaps of the other image. The implied rotation rates are high (90\degr\ in a few days) but this seems perfectly plausible given that the higher SNR rotation event in image~A places an upper limit of 5~d on such a swing.

Alternatively, the PPA of Epoch~37 has a rather large uncertainty and thus it could be that the apparent second rotation event is not real. We have also examined the $u,v$ data and calibration tables very carefully and have been unable to identify any problems with this epoch. However, such events have been observed in other sources, for example in 3C~454.3 where monitoring by the F-GAMMA program \citep{fuhrmann16} using the 100-m Effelsberg telescope at 8.4~GHz has witnessed the same behaviour of a brief return to the pre-rotation state before settling at the post-rotation PPA value \citep{myserlis16,angelakis17} again in combination with a minimum in the polarized flux density. Therefore, we have no real reason to doubt that the apparent second rotation event is real and ascribe the lack of a counterpart in the other image as being due to the sampling rate (0.2~d$^{-1}$ on average) being too low.

Of the two alternatives for the time delay, the lower value (144-146~d) is the most likely. Firstly, this gives a better fit i.e.\ a (marginally) lower $\chi^2$ and a higher cross-correlation. Secondly, the percentage polarization data also favour the lower delay -- the CCF drops sharply beyond 156~d and 161~d corresponds to a local maximum in the CSM distribution.

Further evidence for the lower value comes from a consideration of the polarization ratio, $p_{\mathrm{A/B}}$, between the two images. If not equal to 1, this indicates that some propagation effect is changing the polarization of at least one of the images, the most likely being depolarization in the ISM of the lensing galaxy. The $\chi^2$ distribution for percentage polarization has a global minimum at 156~d for which $p_{\mathrm{A/B}} = 0.72$, thus indicating that such effect is indeed taking place (Table~\ref{tab:delay}). However, A is less polarized than B which is unlikely given that the emission from image B passes much closer than A (0.78 versus 9.74~kpc\footnote{Assuming a flat $\Lambda$CDM cosmology and using Ned Wright's JavaScript Cosmology Calculator \citep{wright06}.}) to the main lensing galaxy. The polarization ratio is a strong function of trial time delay and declines steadily from a value of $\sim$1.2 at 130~d to $\sim$0.5 at 170~d. A ratio of $p_{\mathrm{A/B}} = 1$ (no relative depolarization) corresponds to a delay of 142--143~d.





\begin{figure*}
  \begin{center}
    \includegraphics[scale=0.4]{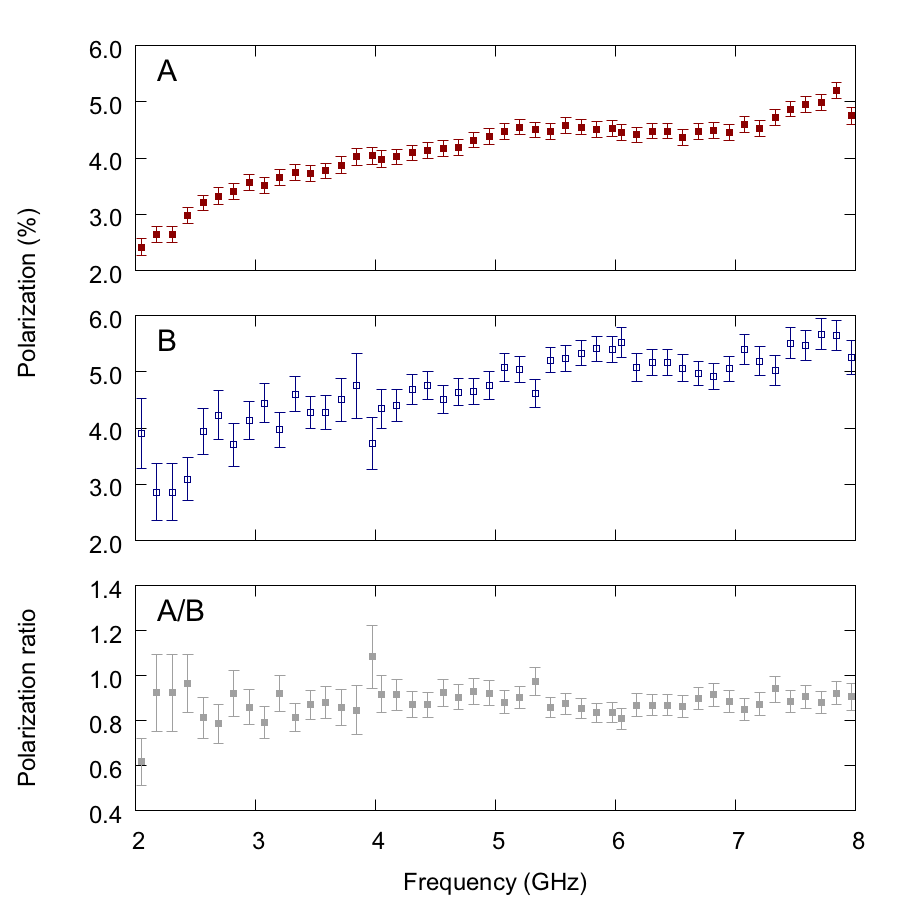}
    \caption{Percentage polarization of images A and B as a function of frequency. Each point corresponds to a 128-MHz-wide subband. The bottom panel shows the ratio of the two images and demonstrates that there is no difference in the polarization dependence of each with frequency. The third subband from the bottom was flagged by the pipeline due to strong radio frequency interference which is more prevalent in the 2--4~GHz frequency band. Source variability together with the different effective epochs of observation (due to the time delay) is most likely responsible for the average polarization ratio being $<1$.}
    \label{fig:mao}
  \end{center}
\end{figure*}

In order to constrain the polarization ratio, we have analysed archival VLA data taken in 2012 (project code 12B-180, PI Sui Mao) which covers the frequency range 2--8~GHz. The data were calibrated using the VLA CASA pipeline with the polarization calibration performed manually using 3C~84 as the leakage calibrator and 3C~138 as the absolute PPA reference. Modelfitting was performed exactly as with the monitoring data. The results are shown in Fig.~\ref{fig:mao} and demonstrate that each image slowly depolarizes with decreasing frequency. Their ratio, however, shows no significant frequency dependence and therefore we conclude that this depolarization is intrinsic to the lensed source and that there is no significant differential depolarization occurring in the lensing galaxy. The frequency-averaged polarization ratio is $<1$, but this will be due to source variability combined with the time delay (we are seeing the images at different epochs).

The final delay estimate and uncertainty is determined using Monte Carlo simulations following the model-dependent procedure described by \citet{biggs18}. This forms a composite variability curve using the best-fit time delay and $y$ offset. A smoothing spline is then fitted to this from which simulated A and B variability curves can be generated using randomized sampling intervals and perturbing each epoch within its error bar. From 5000 realisations of the PPA data using the CSM technique we find a delay of $146\pm6$~d (1-$\sigma$) and a differential rotation (A$-$B) of $1^{+8}_{-7}$\degr. Due to its large error bar, the smoothing spline ignores the discrepant PPA measurement in image B and thus the goodness of fit of the simulated data is better than that found using the real data. Whilst this may reduce the uncertainty found using the Monte Carlo simulations, this is much preferable to having a sharp feature in the data that would most certainly lead to an underestimate of the uncertainty. The same process applied to the percentage polarization data while fixing $p_{\mathrm{A/B}} = 1$ results in a similar value of $147^{+6}_{-4}$~d. To be conservative we choose the PPA result as our final delay estimate as the errors are slightly larger and less assumptions go into its production.


\section{Discussion}
\label{sec:discussion}

\begin{figure*}
  \begin{center}
    \includegraphics[scale=0.27]{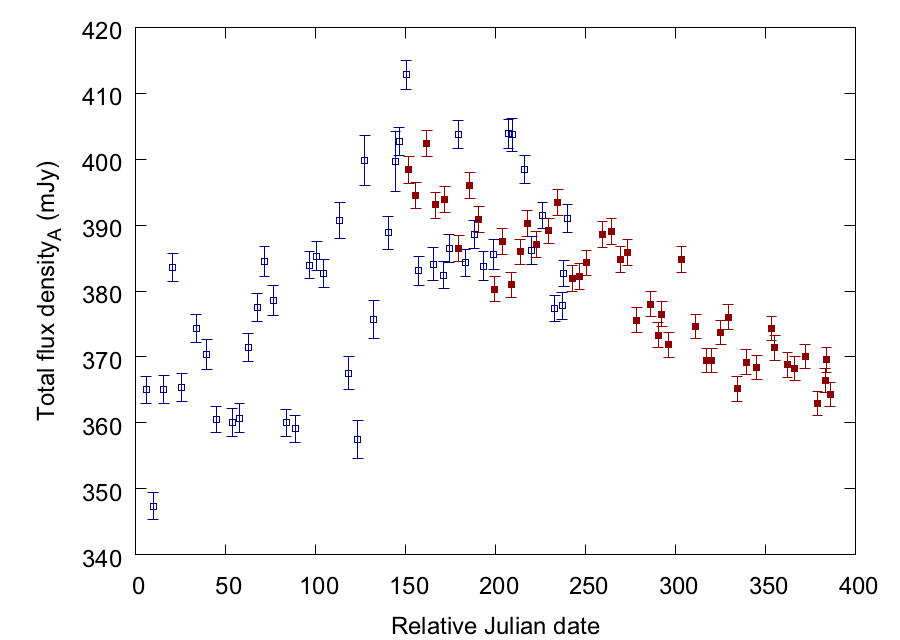}
    \includegraphics[scale=0.27]{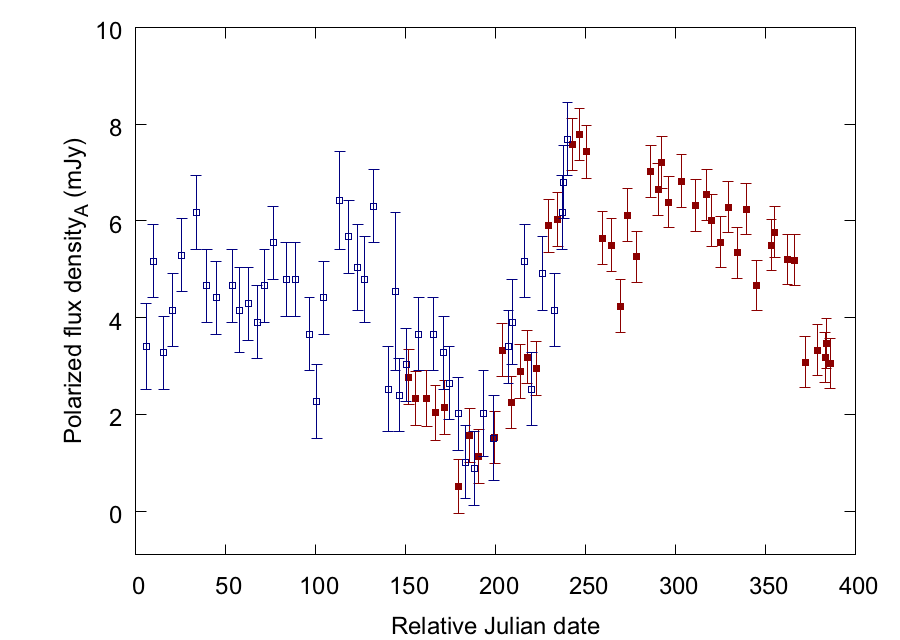}
    \includegraphics[scale=0.27]{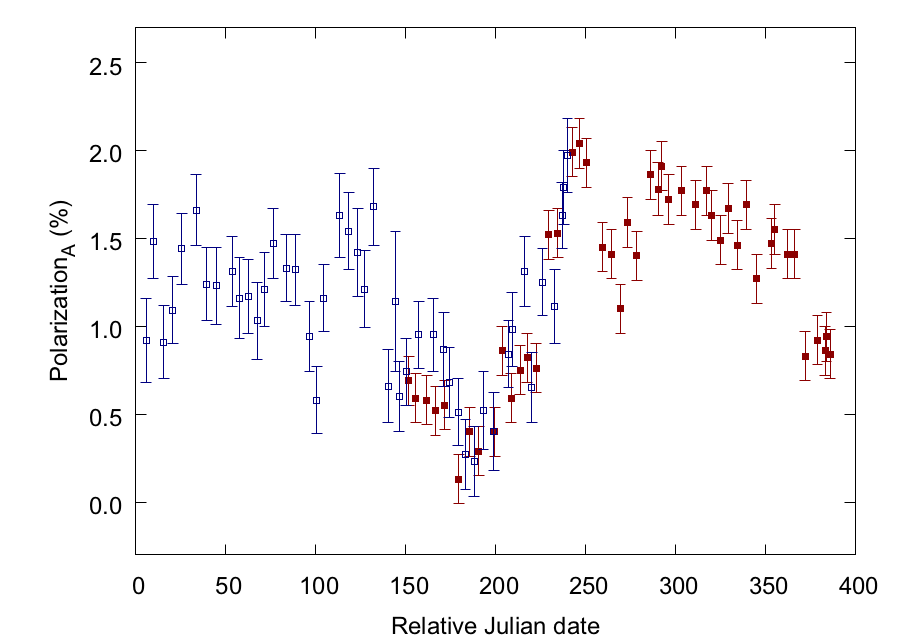}
    \includegraphics[scale=0.27]{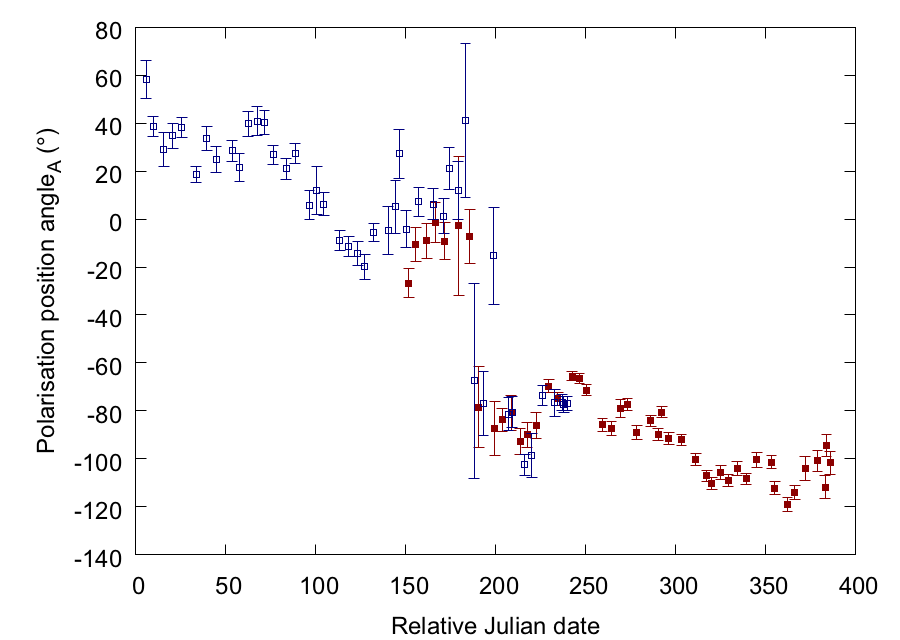}
    \caption{Combined 8.4-GHz variability curves after removal of the time delay (146~d) and $y$ offsets (flux ratio, depolarization or differential rotation). The offsets have always been removed such that B is scaled or rotated to the image A values. Top-left: total flux density, top-right: polarized flux density, bottom-left: percentage polarization and bottom-right: polarization position angle. Image A: filled symbols, image B: open symbols.}
    \label{fig:axlc_comb}
  \end{center}
\end{figure*}

Combined variability curves are shown in Fig.~\ref{fig:axlc_comb} where in each case image A has been shifted forward by our adopted time delay of 146~d. For the PPA data, the differential rotation of 1\degr\ has also been removed. In the case of total flux density, although we do not consider these data capable of measuring a time delay, they do constrain the flux density ratio quite well. For the trial delay range of 110--180~d, the CSM technique finds flux ratios within the range 12.54--12.69 (the value at 146~d is 12.61) and thus we assign a value of $12.6\pm0.1$~d to this parameter and multiply the total and polarized flux densities of image B by this amount, again assuming zero differential depolarization between A and B.

As a check on the assumption that there is no external depolarization, $p_{\mathrm{A/B}} = 1$, we have also run our Monte Carlo simulations on the polarized flux density data. The polarized flux ratio at our measured time delay of 146~d is 11.5, somewhat smaller than that found using the total flux density data. However, simulated variability curves created using this delay and $y$ offset result in a polarization flux ratio of $11.5\pm1.5$, where the error is again 1~$\sigma$. Therefore, the total flux and polarized flux ratios are consistent with each other, supporting the assumption that A and B are depolarized by the same amount at this frequency.

A notable aspect of the total flux density curves is that image B is considerably noisier than A. It is quite clear from Fig.~\ref{fig:axlc_comb} that the two images agree poorly during the overlap period and the reduced $\chi^2$ found using the CSM technique is very high, never falling below 6. The rms around a linear fit to the total intensity variability curve of image B is twice as large as that in A. Such differential effects can only occur in a medium that is external to the lensed source and this is most likely located within one or more of the potentially multiple lensing galaxies which lie close to the line of sight to image B.

A very similar effect has been observed in another radio lens, CLASS~B1600+434 \citep{koopmans98}. VLA monitoring at 8.4~GHz revealed much greater variability in image A of this system \citep{koopmans00a} and this was interpreted by \citet{koopmans00b} as evidence for microlensing by compact objects in the halo of the spiral lensing galaxy. With only a single frequency it is not possible to rule out interstellar scintillation in our Galaxy, but subsequent multi-frequency monitoring \citep{koopmans01} seems to support the microlensing hypothesis. Both explanations require that the source have a small angular size. For B1030+074, the much smaller size of image B (by a factor of $\sim$12.6 relative to A) also makes it more likely that it will be more affected by external variability.

The \citet{rumbaugh15} analysis of the 2000/2001 monitoring of B1030+074 also report excess noise in the difference between A and B and from our own analysis of these data we conclude that image B is indeed varying more than A. The external variability in this lens therefore seems to be a long-lived effect, but without multi-frequency data it is impossible to draw a conclusion as to the cause of the excess variations in B. The high Galactic latitude of this source ($+$52\degr) might favour the microlensing explanation.

Although \citeauthor{rumbaugh15} performed a time-delay analysis of their B1030+074 data, they did not consider that they had been able to measure a reliable time delay and do not quote any possible values. However, their fig.~8 shows quite clearly that, especially for their Monte Carlo Markov Chain $\chi^2$ method, the most likely parameters correspond to a delay of between 140 and 150~d and a flux ratio of $\sim$12.5--12.6. Given the issue of the external variability, the fact that this completely independent analysis agrees so well with our polarization-based values is remarkable and strong evidence that our time delay is correct.

\section{Conclusions}
\label{conclusions}

Analysis of the polarization variability of the gravitational lens system JVAS~B1030+074 using previously published VLA 8.4-GHz monitoring data has allowed us to measure the time delay between images A and B with an accuracy of 4~per~cent ($146\pm6$~d). We have not attempted to measure a delay in total flux density, partly because image B is subject to additional variability that is presumably caused by passage of the radio waves through the lensing galaxy or galaxies. Additional monitoring at multiple frequencies should be able to distinguish between a microlensing or scintillation origin. Despite the difficulties caused by the external variability, an analysis of total-intensity monitoring from 2000/2001 \citep{rumbaugh15} seems to be consistent with our time delay.

In principle, B1030+074 now joins the ranks of those lenses for which $H_0$ can be determined. However, the measured delay is much longer than that predicted by the model published by X98, thus indicating that this is an inaccurate representation of the mass responsible for the lensing. It is not clear which value of the flux density ratio was used as a modelling constraint, but the accurate measurement of this parameter may lead to a more accurate lens model. At the same time, \citet{saha06} use non-parametric modelling to predict a delay that is much closer to our measured value, $153^{+29}_{-57}$~d, albeit with large error bars.

To improve future modelling attempts, more constraints would ideally be needed and one route that in our opinion remains promising is VLBI imaging. It has long been known that image A contains a prominent jet, but the large flux ratio renders this difficult to detect in B as its surface area is reduced by this amount relative to A. Almost all imaging to date has been conducted at 1.4~GHz and we recommend new imaging efforts at 5~GHz. This allows an increase of angular resolution by a factor of three and the dramatic increase in sensitivity offered by modern broad-band ($\Delta\nu = 1$~GHz) VLBI arrays will ensure that the detectable length of the jet will be at least as long as in the existing maps, assuming a standard synchrotron spectrum with a spectral index of $-$0.7.

Finally, the dramatic variations seen in B1030+074 highlight that gravitational lens monitoring campaigns should always attempt to detect the source polarization, even if this is believed to be weak. The SNR of image B in particular is generally very low, occasionally dropping below 3. However, the higher variability seen in the polarization data more than compensates for this and has led to the measurement of a time delay in a lens system for which this had not seemed possible despite two extensive monitoring campaigns. The enhanced sensitivity of the broad-banded Jansky VLA in particular should allow polarization monitoring of more lens systems.

\section*{Acknowledgements}

The author wishes to thank Ian Browne for his many contributions to this project. Much thanks also goes to John Wardle for stimulating discussions regarding 90-degree PPA rotations and the magnitude of polarization position angle errors at low SNR. Finally, we thank the anonymous referee for a careful reading of the manuscript and various recommendations that improved the paper. The National Radio Astronomy Observatory is a facility of the National Science Foundation operated under cooperative agreement by Associated Universities, Inc.




\bibliographystyle{mnras}
\bibliography{lensing}


\bsp	
\label{lastpage}
\end{document}